\documentstyle[aps,preprint]{revtex}
\begin{document}
\title{Instabilities of  Charged Polyampholytes}
\author{Yacov Kantor}
\address{Department of Physics, Massachusetts Institute of  
Technology,
Cambridge, MA 02139, U.S.A.\\
Physics Department, Harvard University, Cambridge MA 02138, U.S.A.\\
School of Physics and Astronomy, Tel Aviv University, 
Tel Aviv 69 978, Israel\cite{permad}
}
\author{Mehran Kardar}
\address{Department of Physics, Massachusetts Institute of  
Technology,
Cambridge, MA 02139, U.S.A.}
\date{\today}
\maketitle
\tightenlines
\begin{abstract}
We consider polymers formed from a (quenched) random sequence 
of charged monomers of opposite signs. Such polymers, known 
as polyampholytes (PAs), are compact when completely neutral 
and expanded when highly charged. We examine the transition 
between the two regimes by Monte Carlo simulations, and by 
analogies to charged drops. We find that the overall excess  
charge, $Q$, is the main determinant of the size of the PA. 
A polymer composed of $N$ charges of $\pm q_0$ is compact 
for $Q<Q_c\approx q_0 \sqrt{N}$, and expanded otherwise. The 
transition is reminiscent of the Rayleigh shape instability 
of a charged drop. A uniform excess charge causes the breakup 
of a fluid drop, and stretches out a polymer to a 
{\it necklace} shape. The inhomogeneities in charge distort 
the shape away from an ordered necklace.
\end{abstract}
\section{Introduction}  \label{secintro}
Polyampholytes (PAs) are long chain macromolecules with 
a random  mixture of oppositely charged groups fixed along 
their backbone\cite{rPA}.  Several elements conspire to 
make the behavior of such heteropolymers a problem of great
interest. One (admittedly somewhat remote) motivation is the  
similarity to the macromolecules of biological interest such 
as nucleic or amino  acids. The specific sequence of 
monomers on such chains is essential to biological activity. 
For example, the sequence of amino acids  determines  the 
ultimate shape of a protein\cite{rgen}. Attempting to unravel 
the  precise factors responsible for {\it protein folding}, 
several statistical  models have been proposed\cite{rstein}. 
These models sacrifice the specifity of particular proteins, 
by essentially focusing on generic properties of
heteropolymers with competing interactions\cite{rGOa}. PAs 
can be regarded as a particular example of this class. From 
another perspective, properties of random systems with 
competing interactions has been on  the forefront of 
statistical mechanics for the past  decade\cite{spinglass}.
The prototype of complexity in this class of problem, the 
spin glass, has much in common with random heteropolymers. 
The statistics of the ground state, and those of low lying 
excitations, is paramount to both  systems. As examples 
of {\it soft} condensed matter, heteropolymers have the
advantage of faster relaxation, compared to their ``harder"  
counterparts. Indeed there is much encouragement from recent 
experimental studies of solutions\cite{rcand} and gels 
composed of  PAs\cite{rExT,anaka}.

It may appear that the long range nature of the Coulomb 
interaction between charges is yet an additional complication 
of an already hard problem. Yet for a uniformly charged a 
polymer (a polyelectrolyte), it is  possible to find the 
{\it exact} scaling of the radius of gyration $R_g$ on  
the number of monomers $N$\cite{rPVdG}. The proof relies 
on the non-renormalization of the dimensionless interaction 
parameter $u=Q^2/(k_BTR_g^{d-2})$ in $d$ embedding dimensions. 
Since $Q\propto N$, it follows that $R_g\sim N^\nu$, with 
$\nu=\nu_h\equiv 2/(d-2)$ (for $4\leq d\leq 6$). Inspired by 
the simplicity of the homogeneous case, we suggested a similar 
argument for the randomly charged PA\cite{rKK}. Consider a  
model PA composed of $N$ charges  $\pm q_0$, randomly and 
independently chosen at each site. Although the mean net  
charge is zero, a typical PA has excess charge of order 
$\pm q_0\sqrt{N}$. Independent of its sign, this leads to 
a repulsive self energy, on average described by the 
dimensionless parameter 
$\overline{u}=Nq_0^2/(k_BTR_g^{d-2})$.
Assuming that this parameter is not renormalized as in 
the uniformly charged case leads to a swelling exponent of 
$\nu=1/(d-2)$ (for $3\leq d\leq4$), i.e. a polymer that 
is {\it stretched} in $d=3$.

However, in an electrolytic solution oppositely charged
ions rearrange so as to screen the long range Coulomb 
interaction. The net effect is an attractive 
energy\cite{landau}, described by the Debye--H\"uckel (DH) 
theory. Higgs and Joanny\cite{rHJ}, assumed that the 
monomers in a PA can similarly rearrange to {\it compact}
configurations, thereby taking advantage of the DH attraction.
A partial resolution of the contradiction between the two 
predictions is obtained by noting that DH theory requires 
the exact neutrality of the electrolyte, while the  
RG--inspired approach depends on the excess charge in a 
typical sequence. Monte Carlo 
simulations\cite{rKLKprl,rKKLpre} indeed confirm that PAs
with $Q=0$ compactify at low temperatures. By contrast 
sampling all random quenches with unrestricted $Q$  
produces a broad range of sizes, with an average consistent 
with $\overline{R_g}\propto N$.

Depending on the conditions, experiments observe both 
compact and expanded conformations\cite{rcand}. One set of 
experiments\cite{rExT} were performed on gels produced by 
crosslinking PAs\cite{anaka}. By changing the conditions of 
the solvent (pH, salt content, etc.), it is possible to 
control both the excess charge $Q$ on the PA chains, and
the screening length. Due to the large screening length, the 
Coulomb interactions are significant. As a function of $Q$, 
the gel undergoes dramatic changes in volume; the neutral 
gel being the most compact. However, the volume of the gel 
does not change gradually with increasing charge. There is 
an  interval of $Q$ around the neutral point where the gel
remains compact, suddenly increasing in volume by a order 
of magnitude beyond a threshold $Q_c$. The threshold scales 
with the number of monomers $N$, within a screening
length as $Q_c\approx q_0 \sqrt{N}$. Motivated by the 
experimental results, we undertook a systematic 
examination of the dependence of the size of a PA on its 
excess charge. Monte Carlo simulations show that PAs are 
compact for small $Q$,  and  expanded when $Q$ exceeds a 
critical value of  $Q_c\approx q_0\sqrt{N}$, in complete 
analogy with the experiments. Some aspects of this 
transition can be understood by analogy to the behavior 
of a charged drop. The spherical drop is stable for 
charges smaller than a Rayleigh limit $Q_R\propto\sqrt{V}$,  
where $V$ is the volume of the drop. Charged beyond this 
point, the drop elongates to minimize the Coulomb 
repulsion. The elongated drop rapidly disintegrates into 
smaller droplets. In attempting to follow a similar 
scenario, a PA chain breaks into a necklace of globules
connected by strings. The detailed shape of the necklace 
is determined by charge inhomogeneities.

The main results of this work were summarized in an 
earlier publication\cite{KKshort}. Here we provide more 
detailed results and additional information. The paper 
is organized as follows: The competing arguments applied 
to PAs are discussed in some detail in Section 
\ref{secgen}, and their inconsistencies are emphasized.
The results of Monte Carlo simulations on the dependence 
of the radius of gyration of the polymer on temperature 
and excess charge are presented in Section \ref{secTdep}. 
Various details of the Monte Carlo procedure are relegated 
to Appendix \ref{MC}. We argue that as $Q$ is increased 
beyond a threshold $Q_c$, the PA undergoes a sudden 
transition from a compact to a strongly elongated state.
In Section \ref{secQdep} we provide a qualitative picture 
of this transition by analogy to the shape instability 
of a charged drop. Some known results pertaining to such 
drops, as well as new calculations for spheroidal shapes
are presented in Appendix \ref{secchdrop}. Such analogies 
cannot be extended to the strongly distorted limit where, 
as argued in Section \ref{secstrong}, a uniformly charged 
polymer deforms into a necklace of compact beads. Such a 
shape is the best compromise for the PA in trying to 
mimic the ground state of a charged drop, which as 
described in the Appendix \ref{secground}, is obtained by 
splitting into several droplets. In Section 
\ref{secbeyond} we point out the importance of quenched  
randomness in the PA. An ordered necklace is not stable 
to charge inhomogeneities, and the beads must rearrange 
in complicated shapes dependent on the details of 
randomness.

\section{Polyampholyte Phenomenology}\label{secgen}

In this work we consider PAs immersed in a good solvent 
in which the concentration of counterions is small, and 
hence the electrostatic interactions are treated as 
unscreened. Experimentally, the details of a charge 
sequence are determined by its fabrication process.
{\it Markovian} sequences are constructed by adding
one monomer at a time, the probability of choosing a
particular monomer (e.g. positively or negatively
charged) depends only on the last monomer\cite{wittmer}.  
Such a construction leads to correlations in the charges 
$q_i$, which decay exponentially as,
\begin{equation}
\overline{q_iq_j}=q_0^2\lambda^{|i-j|},\quad -1<\lambda<1.
\end{equation}
The extreme limits of $\lambda=-1$ and $\lambda=1$ correspond
to {\it non-random} sequences that are 
alternating\cite{victor}, or fully charged,
and will not be considered here. The behavior for $\lambda$ is
asymptotically similar to the case where the charges are
uncorrelated ($\lambda=0$), and we shall focus mostly on such chains.

We shall, however, allow for the possibility that the total charge
of the chain is constrained to a particular value of $Q$. 
Experimentally, this net charge can be controlled by
changing the pH, and other properties of the solvent.
At several points we  shall also contrast the behavior of 
{\it quenched} and {\it annealed}  PAs; 
the latter is defined a predetermined  
collection of positive and negative charges which can 
freely move along the polymer chain. Note that in this 
definition the total charge is fixed. If this constraint 
is also removed the polymer lowers its energy by  
getting rid of any excess charge.
Note that such annealed PAs are not very realistic,
and should not be confused with the case considered
by Raphael and Joanny\cite{raphael}, where the excess charge
can adjust its value depending on the temperature, 
concentration of PAs and concentration of the counterions.

A short $\ell$--monomer segment of  a PA with uncorrelated
random monomers has typical charge of  $q_0\sqrt{\ell}$.
If we assume that the segment is a self--avoiding
walk, its radius of gyration is approximately 
$a\ell^{\nu}$, where $a$ is a  microscopic length 
(e.g. the monomer diameter or nearest neighbor separation 
along the chain). As the typical electrostatic energy of 
such a subchain is $q_0^2\ell^{1-\nu}/a$, interactions 
become important for $T\approx q_0^2\ell^{1-\nu}/a$. (We 
shall henceforth measure temperature in energy units, 
i.e. set $k_B=1$.) Alternatively, we can define 
$\ell_T=(Ta/ q_0^2)^{1/(1-\nu)}$, and divide the entire 
chain into segments of $\ell_T$ monomers. The 
interactions within each segment are small compared with 
$T$, while interactions between the segments are strong.
Such segments form the basic ``ions" in an 
analogy\cite{edwards} between PAs and usual 
electrolytes\cite{landau} employed by Higgs and
Joanny\cite{rHJ}. The spatial extent of each segment is
\begin{equation}
a_T=a\ell_T^\nu=\cases{a\left({Ta\over q_0^2}\right)
^{\nu/(1-\nu)} &,  for $T>q_0^2/a$,\cr
                a &, for $T<q_0^2/a$.\cr}
\end{equation}

For the generalized Markovian chains, the net charge of a
sufficiently long segment (such that $\lambda^\ell\ll1$)
grows as $q_0\sqrt{N(1+\lambda)/(1-\lambda)}$. The
previous argument thus remains valid after replacing
$q_0$ by $q_0\sqrt{(1+\lambda)/(1-\lambda)}$. This
simple change is sufficient to relate most high temperature
properties of the short-range correlated sequences to
the uncorrelated ones (e.g. in a high  temperature series
expansion for the radius).

In the spirit of an RG analysis, we can attempt to increase 
the short  distance cutoff  along the sequence by a factor $\ell$. 
Perturbations around the high temperature phase depend 
on the {\it dimensionless} interaction  parameter
$u\equiv q_\ell^2/r_lT\ll1$, where $q_\ell$ is a typical 
charge on length--scale $\ell$, while $r_\ell$ is the 
spatial extent of the shortest segments. Upon rescaling the
cutoff by a factor $b$, $q_\ell$ increases by $\sqrt{b}$, 
while $r_\ell$ scales by $b^\nu$, as in a self--avoiding 
walk. Thus, the renormalized interaction parameter,
$u(b)=b^{1-\nu}u$ grows under rescaling and reaches unity 
for $\ell$ equal to $\ell_T$ defined above. At this scale 
$r_\ell$ becomes of the same order as $a_T$. Beyond this point the 
interactions are {\it relevant}, strongly modifying the 
behavior of  the chain.

Different approaches to the problem are in agreement in 
the weak coupling regime of $u\ll1$. The strong coupling 
regime is not easily tractable, and different assumptions 
lead to different conclusions. Higgs and Joanny\cite{rHJ} 
construct the free energy of a long PA by first assuming 
that it has a uniform density and then proving this 
assumption self-consistently. (See also 
Refs.\ \cite{rKLKprl,rKKLpre}.) In this approach the 
$N$--monomer PA is divided into blobs\cite{scaling} of 
$\ell_T$ monomers each, forming a liquid of uniform density 
as depicted  qualitatively in  Fig.\ \ref{FigO}a. The blobs 
are non--interpenetrating and arranged so that the  
neighborhood of each blob is predominantly occupied by 
blobs of opposite charge. This arrangement roughly
resembles the structure of a salt crystal. The excess charge 
of each  blob is effectively screened and the dense 
configurations take energetic advantage of the large number 
of neighbors of opposite sign. The energy gain per blob
is approximately the nearest neighbor interaction, i.e.
$\epsilon_c(T)\approx q_0^2\ell_T/a_T$. We note, however, 
that even in the ideal {\tt NaCl} crystal, the condensation 
energy per atom ($0.874q_0^2/a$) is a small fraction of the 
interaction energy between the nearest neighbors 
($3q_0^2/a$), and is thus strongly influenced by  further
neighbors. The validity of the picture depicted in 
Fig.\ \ref{FigO}a rests on the assumption that it is 
possible to fold a randomly  charged object in a way that 
not only provides the correct neighborhood to  each
charge, but also keeps more extended neighborhoods 
approximately neutral. Thus such configurations require 
the possibility of specific foldings of the PA  at both
local and global levels.

The primary focus of the DH--type approach is the 
minimization of the extensive part of the energy by 
creating a homogeneous liquid--like structure, while 
non--extensive energies due to surface tension and electrostatics 
are relegated the role of determining the overall 
shape of the globule.  By contrast, an RG--inspired 
approach\cite{rKK} to the problem assumes that
the blobs form a {\it self--similar} structure (as 
depicted in Fig.\ \ref{FigO}b) which attempts to take 
care of energies on every length--scale. In $d$ space 
dimensions, the dimensionless interaction parameter
at scale $\ell$ is $u(\ell)=q_0^2\ell/(Tr_\ell^{d-2})$. 
This expression represents the typical interaction 
energy of a random $\ell$--monomer segment, 
{\it assuming} that the Coulomb interactions can not 
be screened. For a self--similar structure with 
$r_\ell\propto \ell^{\nu'}$,  $u(\ell)\propto
\ell^{1-(d-2)\nu'}$, and the interaction parameter 
grows or diminishes under rescaling unless 
$\nu'=1/(d-2)$. (The analogous argument leads to the  
exact value of $\nu_{h}=2/(d-2)$ for uniform 
polyelectrolytes where no screening is 
possible\cite{scaling}.) This result is valid only 
for $3\le d\le4$: For $d>4$ the electrostatic
interactions are irrelevant and $\nu=1/2$, while for 
$d<3$ the polymer is stretched ($\nu'=1$). Keeping 
interactions equally strong on all length--scales also
generates a condensation energy propotional to $Nq_0^2/a$. 
However, as this argument does not provide the 
prefactor, i.e. the actual value of the condensation 
energy, it is not possible to deduce whether the 
DH--type or RG--type ansatz produces the lower free 
energy state.

Are the two aforementioned approaches mutually 
exclusive? In this work we present evidence that they 
actually represent two facets of the same problem: 
DH--theory attempts to minimize the condensation 
energy without paying attention to the surface. 
However, an object can be (locally) ``compact'' and 
still have an extremely extended shape which is 
controlled by the non--extensive part of the energy. 
The RG--inspired approach attempts to accommodate the 
latter energy.

\section{Numerical Simulations}  \label{secTdep}

Configurations of a polymer are completely specified by 
listing the position verctors $\{{\bf r}_i\}$  
($i=1,\dots,N$) of its monomers. The shape and spatial 
extent of the polymer are  roughly characterized  by
the shape tensor,
\begin{equation}\label{shapetensor}
{\cal S}_{\mu\nu}={1\over N}\sum_{i=1}^Nr_{i\mu}r_{i\nu}
-{1\over N^2}\sum_{i=1}^Nr_{i\mu}\sum_{j=1}^Nr_{j\nu}\ ,
\end{equation}
where the greek indices denote the Cartesian components 
of the vectors. Thermal averages of the ordered eigenvalues 
$\lambda_1>\lambda_2>\lambda_3$ of this tensor
(sometimes referred to as moments of inertia) are used to
describe the mean size and shape; their sum, i.e. the 
thermal average of ${\rm tr}{\cal S}$, is the squared 
radius of gyration $R_g^2$. Since we are dealing with  
sequences of quenched disorder,  these quantities must 
also be averaged over different realizations. In three 
dimensions, uniform uncharged polymers in good solvents 
are swollen; their  $R_g$ scaling as  $N^\nu$ with 
$\nu=0.588$ as in self--avoiding walks. Polymers in poor 
solvents are ``compact'', i.e. described by $\nu={1\over3}$.

The Monte Carlo procedure used in this work is 
identical to that described in Ref.\cite{rKLKprl}. 
Here we describe some important features, while the 
more technical details are described in Appendix 
\ref{MC}. The simulated chains are composed of
$N$ monomers whose positions are discretized to a 
cubic lattice ($d=3$) with lattice constant $a$.
The connectivity of the polymer is maintained by 
restricting the maximal distance between neighbors 
to $4a$.  The excluded volume interaction is 
enforced by not allowing two monomers to come any 
closer than $\sqrt{2}a$. Each quench is 
characterized by a set of charges $q_i=\pm q_0$. 
The electrostatic interactions between the charges,
${\cal U}=\sum_{\langle i,j\rangle} 
U_{ij}(|{\bf r}_i-{\bf r}_j|)$,
is included  by assigning energy 
$U_{ij}(r)=q_iq_j/\sqrt{c+r^2}$
to each pair ${\langle i,j\rangle}$ at a separation 
distance $r$, with $c=2a^2$, which ``softens'' the
potential at short distances.

The results of the simulations are parametrized by 
the chain length $N$, temperature $T$, and the 
overall excess charge $Q=\sum_i q_i$. Each $Q$ 
can be obtained by many realizations of randomness, 
and all results were averaged over 10 different quenches. 
However, rather than taking the same configurations 
through changing temperatures, 10 distinct quenches  
were used for each $T$ and $Q$. The smoothness in 
variations of various quantities with temperature then 
provides added confidence in the thermal and quench
averaging process. Not surprisingly, as explained in 
Appendix \ref{MC}, the overall uncertainties are 
entirely due to quench averaging as the statistical
errors of the thermal averages are smaller than the 
differences between quenches.

Fig.\ \ref{FigA} depicts the temperature dependence of 
$R_g^2$ for 64--monomer chains. The number near each 
curve indicates the charge, $Q/q_0$. At very high 
temperatures the electrostatic interactions are 
unimportant and the chains behave as self--avoiding 
walks, with $R_g\propto N^\nu$ and $\nu=0.588$.
The typical electrostatic energy of such 
configurations is estimated as
\begin{equation}
\overline{\langle U\rangle}\approx \sum_{
i,j}\overline{q_iq_j}\left\langle{1\over
 |{\bf r}_i-{\bf r}_j| }\right\rangle\approx
 {(Q^2-q_0^2N)\over R_g} ,
\end{equation}
where we have employed 
$\overline{q_iq_j}=(Q^2-q_0^2N)/N^2$ for $i\ne j$, 
and used $R_g$ as a measure of interparticle 
separation. Note that the interaction changes sign 
at 
\begin{equation}
Q_c= q_0\sqrt{N}\ .
\end{equation}
This is because the energy of strongly charged polymers 
is dominated by the {\it repulsive} interaction of excess 
charges. However, for weakly charged polymers, there is 
an {\it attractive} interaction between fluctuations in 
the charge distribution; the typical fluctuation of $Q$ 
leads to the above result.

As temperature of the chain is lowered, the effects of 
interactions become apparent for
\begin{equation}
T_Q\approx\overline{\langle U\rangle}\approx (Q^2-q_0^2N)/
aN^{\nu} \  . 
\end{equation}
Chains with charge larger than $Q_c$ expand, while 
those with $Q<Q_c$ shrink with decreasing temperature.
The strongest deviation occurs for the fully charged 
polymer, which for $N=64$ starts at $T_Q\approx320q_0^2/a$. 
For $Q=0$, the deviation begins
at the much lower temperature of $T_Q\approx 5q_0^2/a$.
Indeed, on the logarithmic scale of Fig. \ref{FigA} the 
departure from the infinite temperature values of $R_g^2$ 
is most apparent for $Q=64q_0$, starting at 
$\ln(T)\approx 6$ (beyond the limits of the figure), compared to
$\ln(T)\approx 1.7$ for $Q=0$.

To see if the {\it averages} in Fig.\ \ref{FigA} provide a 
good measure of the PA size at low temperatures, we 
constructed histograms of the distribution of $R_g^2$
at $T=0.05q_0^2/a$ for several values of $Q$. As 
thermal fluctuations are small, the histograms 
in Fig.\ \ref{FigB} represent differences between quenches. 
The distributions are fairly narrow, their widths 
not exceeding the distance between their averages. Thus a 
point in Fig.\ \ref{FigA} provides a good measure of 
$R_g^2(Q)$, independent of further details of the sequence.
The average of $R_g^2$ for {\it unrestricted} quenches
is then obtained from $R_g^2(Q)$ as,
\begin{equation}
R_g^2({\rm random})=\int_0^\infty dQ R_g^2(Q) P(Q)\ ,
\end{equation}
where
$P(Q)\propto\exp[-Q^2/(2q_0^2N)]$ is the probability 
density of an excess charge $Q$. In previous 
work\cite{rKLKprl}, we found a very  broad distribution for 
$R_g^2({\rm random})$. Even in sampling a few quenches, 
there were several completely collapsed, and some strongly 
stretched configurations. Fig.\ \ref{FigF} shows the spatial 
conformations of several quenches examined in this study. 
The weakly charged configurations for $Q/q_0=0$ or 4 are  
spherical globules,  indistinguishable from each other. 
The chains are slightly expanded for $Q/q_0=8$, while for 
$Q/q_0=16$, a value of only twice $Q_c$, they are strongly 
stretched. It can now be appreciated that the previously 
observed breadth of the distribution for $R_g^2({\rm random})$ 
simply follows from the strong dependence of $R_g^2$ on $Q$
rather than indicating a large scatter of $R_g^2$ amongst 
different quenches with the same $Q$.

The averaged radii of PAs in Fig.\ \ref{FigA} change 
monotonically with temperature. This suggests that compact 
and extended states are separated in the $(Q,T)$ plane by a 
straight line starting from $Q=Q_c=q_0\sqrt{N}$ at infinite 
$T$. This hypothesis was tested by looking at  the $Q$ and 
$N$ dependence of the radius of gyration for chains of 
lengths $N=16$, 32, 64, 128. To achieve good thermal 
averages, simulations were performed at $T=0.1q_0^2/a$ and 
not at the lowest temperature in Fig.\ \ref{FigA}. The 
dependence of $R_g^2$ on $Q$ is depicted in Fig.\ \ref{FigC}. 
The vertical axis is scaled by $N^{2/3}$ to remove the 
$N$--dependence of the $R_g^2$ of the compact globules at  
$Q=0$.  The charges on the 
horizontal axis are scaled by $Q_c(N)$ for all polymer 
lengths. Although monotonic, $R_g^2$ exhibits strong  
variations with $Q$. The radius is barely increasing for 
small $Q$, but an extremely steep rise begins beyond a 
threshold charge. Due to the monotonic increase or decrease of  
PA sizes with temperature, this variation becomes even 
sharper at low temperatures. Fig.\ \ref{FigC} strongly 
suggests that the transition from compact to stretched 
configurations at low temperatures still occurs for 
$Q\approx Q_c$.

When the distortions of a globular object are small, the 
changes in $R_g^2$ are not very sensitive to the changes 
in shape. The increase in the largest eigenvalue of 
the shape tensor $\lambda_1$ due to elongation of the 
object is partially compensated by a decrease of other 
eigenvalues. Fig.\ \ref{FigR} is analogous to 
Fig.\ \ref{FigC} except that it depicts the charge 
dependence of the $\lambda_1$. The strong elongation of 
the PA for $Q>Q_c$ is also apparent from this figure.

\section{Analogies to Charged Drops}\label{secQdep}

To explain the above results, we start with the empirical 
observation that 
PAs with  vanishing excess charge $Q$ compactify to
spherical ``globules'' of spatial extent 
$R\approx a N^{1/3}$ and surface area  
$S\approx a^2N^{2/3}$. It is thus natural to represent
the total energy (or rather the quench--averaged free 
energy) of such globules as a sum of condensation and 
surface energies,
\begin{equation}\label{EQ0}
E_{\rm PA}(Q=0)=-\epsilon_c N+ \gamma S\ .
\end{equation}
The condensation energy is proportional to $q_0^2/a$, while the 
surface tension is $\gamma=pq_0^2/a^3$, where the 
dimensionless prefactor $p\approx0.1$, is found to be rather 
small\cite{KKfuture}.  It should be emphasized  
that Eq.(\ref{EQ0}) {\it is not self--evident} as it 
represents the average energy of a 
{\it connected} chain of $N$ monomers with long range
interactions, rather than $N$ independent particles. 
While the existence of an extensive condensation energy 
is natural, non--extensive corrections may in principle 
be present without any relation to the surface. 
The presense of the surface term is deduced from the 
numerical observations that the object is approximately  
spherical. We may hope that this form of the energy 
persists as long as the deformations of the globule are 
not too large.

If we now uniformly add a very small number ($Q/q_0\ll N^{1/3}$) 
of charges along the chain (e.g. by randomly 
replacing $Q/2q_0$ of negative charges by positive ones), 
without modifying the spatial conformation of the 
PA,  its total energy increases approximately by
$q_0|Q|/a$, representing the sum of changes in {\it local}  
interactions. For moderate charges ($Q>q_0N^{1/3}$) the 
energy increase is dominated by the long range 
interactions and is of order $Q^2/R$. If the PA is now
allowed to relax, it  will lower its energy by finding 
more favorable configurations, and thus, 
\begin{equation}\label{EPA}
E_{\rm PA}(Q)\leq -\epsilon_cN+\gamma S+Q^2/R\ .
\end{equation}
The above considerations are equally applicable to
a charged drop, and we shall explore such analogies to
treat  weakly distorted PAs.  In the following 
paragraphs, we initially review the results  
pertaining to the shape of a charged {\it conducting} 
drop. This analogy is most appropriate for an annealed 
version of the problem in which the charges are free 
to move along the polymer chain. We then go on to 
consider the shape of a charged {\it insulating} drop
of immobile charges, which is a better representation 
of quenched PAs.

For a {\it conducting} drop all the charge accumulates 
on the surface. The non--extensive contribution to the 
energy  of a freely suspended spherical drop of radius 
$R$  is
\begin{equation}
E(Q)={1\over2} {Q^2\over R}+4\pi R^2\gamma\ .
\end{equation}
However, the drop can change shape to minimize the sum 
of surface tension and Coulomb energies.  The surface 
energy of an uncharged drop, $E(0)=4\pi R^2\gamma$, 
sets the overall energy scale of the problem while 
the dimensionless parameter, 
\begin{equation}\label{Esphere}
\alpha\equiv Q^2/(16\pi R^3\gamma) \equiv Q^2/Q^2_R\ ,
\end{equation}
determines its shape. We shall refer to $Q_R$ as the 
{\it Rayleigh charge} of the drop. Note that the 
estimates of $\gamma$ and $R$ following Eq.\ (\ref{EQ0}) 
for the model PA lead to $Q_R\approx Q_c$. Let us 
initially consider only small deformations in shape of 
a single drop. Investigations of the shape of charged 
liquid drops go back at least to the last 
century\cite{Rayleigh}. Some 
of the several variants of this problem are:
(a) a freely suspended charged conducting
drop\cite{Rayleigh,Taylor,Ailam,Hendricks,Grig,Grigor};
(b) an uncharged conducting drop in an external
electric field\cite{Shir,Miksis};
(c) a dielectric drop embedded in a different dielectric 
liquid in an external electric 
field\cite{Sherwood,Brazier,Bacri,Li,Sero}. The last 
problem has 
received recent attention because it is mathematically
identical to that of a drop of  magnetic fluid in a 
magnetic field\cite{Bacri,Sero,BSM,Liao,Hayes}. The 
conducting drop is the limiting case of an insulating 
drop with infinite dielectric constant. Problems (a) and 
(b) are strongly related since similar shape 
instabilities are induced by both the external field,  
and by the internally  generated field of a charged
drop. In fact, in many experimental situations (see, 
e.g. Ref.\cite{Zeleny}), the drop is suspended from a tube. 
Creating instabilities by raising the potential of the 
tube is intermediate  between the idealized situations 
described by (a) and (b).  The stability conditions in
experiments are usually discussed from the point of  
view\cite{Taylor} of increasing potential rather than 
charge. Therefore, many features of the  problem which
are of interest to our study are usually not addressed. 
In the Appendix \ref{secchdrop} we summarize the 
relevant aspects from our perspective, while only  
quoting the main results in this Section.

A closely related problem is that of an {\it insulating} 
uniformly charged drop. This problem has been considered 
in the framework of the charged drop model of atomic 
nuclei (see, e.g., Ref.\cite{weisskopf}). The 
non--extensive portion of the energy of a spherical shape 
is now
\begin{equation}\label{EiQ}
E^{(i)}(Q)={3\over5}{Q^2\over R}+ 4\pi\gamma R^2\ .
\end{equation}
The similarity between Eqs. (\ref{Esphere}) and 
(\ref{EiQ}) is  evident. As explained in the 
Appendix \ref{secchdrop} this similarity persists  
even for the non--spherical shapes discussed in this 
Section: all results for conducting drops are 
transformed into results for insulating drops by 
replacing $Q^2/2$ with $3Q^2/5$. This analogy, 
however, does no persist to arbitrary deformations and
differences between conducting and insulating models 
will become apparent in Section \ref{secstrong}.

If the only allowed deformations of the charged drop 
are to ellipsoids of rotation (prolate spheroids), the 
spherical shape remains stable until $\alpha$ reaches 
0.899. At this point the drop becomes strongly 
elongated with eccentricity $e=0.95$, and continues to 
stretch with increasing $\alpha$. For large $\alpha$
the long axis of the drop (and hence  $R_g$) is 
proportional to $\alpha^{1/3}$. Due to the sharp 
increase in the aspect ratio of the spheroid with  
increasing $\alpha$, the increase in the energy of 
the system slows down and becomes of order of 
$(\alpha\ln\alpha)^{1/3}$ (compared to order of 
$\alpha$ for the undistorted sphere). Fig. \ref{FigL} 
depicts the resulting dependences of $R_g^2$ and $E$  
on $\alpha$. The behavior of $R_g^2$ in 
Fig. \ref{FigL} closely resembles the sudden 
expansion of polyamphilic gels in ref.\cite{rExT}.

A more quantitative comparison between our results 
and the predictions of the charged drop model is 
possible: The transition in Fig.\ \ref{FigC} appears at
$\alpha'\equiv Q^2/(q_0^2N)\approx1$. The instability 
of a charged drop to ellipsoidal shape occurs for
$\alpha\approx0.9$ in the conducting case, and 
$\alpha\approx0.7$ in the uniformly charged case.
In Eq.\ \ref{Ealpha} we show that 
$\alpha=Q^2/(12V\gamma)$, where $V=a^3N$ is the 
volume of the system, while 
$\gamma\approx0.1q_0^2/a^3$\cite{KKfuture}. Thus,
$\alpha\approx\alpha'$, and the observed critical 
value of $\alpha'$ is surprisingly close to the 
predictions of the model. (Given the numerous 
approximations of the model, such excellent 
agreement is probably fortuitous.) We conclude 
that, as long as the value of $\alpha$ is not too 
large, our MC results, the predictions of the 
spheroidal drop model, and  experiments, are in good 
agreement.

For comparison with Fig.\ \ref{FigL}, Fig.\ \ref{FigS} 
depicts the energy per monomer as a function of scaled 
excess charge, obtained from MC simulations. At $Q=0$, 
the curves for different $N$s almost coincide. (Slight 
corrections of order $N^{-1/3}$ are present, but  
invisible at this scale.) For small $Q$, the energy 
per monomer increases as $Q^2/(RN)\sim Q^2/N^{4/3}$, 
as indicated by the straight dashed line in 
Fig.\ \ref{FigS}. For large charges, the energy 
increase slows down, indicating distorted PAs. 
Since distortions start for $Q^2/q_0^2N\approx 1$, in
terms of variable used in  Fig.\ \ref{FigS} the 
departures of longer chains   begin earlier.

The results of this Section are not sensitive to 
the exact shape of the elongated drop. Any shape 
characterized by a long dimension $R_\parallel$, 
and a short dimension $R_\perp$, (such as major 
and minor semi--axes of an ellipsoid), reproduces 
the same answers qualitatively. The electrostatic 
energy is approximately $\sim Q^2/R_\parallel$, 
while surface energy grows as $\gamma R_\perp
R_\parallel$. Their sum has to be minimized 
subject to the constraint of fixed volume,  
imposed by requiring  $V\approx R_\perp^2R_\parallel$.
For example, Gutin and Shakhnovich\cite{rGSpp} 
consider the more general case of $Q\sim N^\beta$. 
Minimizing the total energy for elongated  shapes,
they find $R_\parallel\sim N^{(4\beta-1)/3}$, and 
$R_\perp\sim N^{2(1-\beta)/3}$. Only a finite 
stretching is predicted for $\beta={1/2}$. 
(A directed version of this problem also exhibits
a continuously varying exponent  $\nu(\beta)$\cite{derridahiggs}.)
Another  recent study by Dobrynin and Rubinstein\cite{rDR}
relaxes the constant volume constraint and also 
reaches the conclusion that there is an onset of 
stretching for $\beta={1/2}$, although a 
completely stretched state is reached only for 
$\beta={2/3}$, when the Coulomb energy becomes 
extensive.

However, as we shall show in the following sections, 
the ground state of a charged drop {\it is not} a 
simple elongated shape for large values of $\alpha$. 
Conducting drops can shed away excess charge,
while insulating drops disintegrate in a process 
similar to nuclear fission. Related pathways are 
available to PAs.

\section{Beyond Single Drops}\label{secstrong}

Linear stability analysis indicates that a spherical shape 
is {\it unstable} to a variety of perturbations\cite{Grig}. 
Experiments show that a conducting drop disintegrates when 
the Rayleigh stability limit is exceeded. Nuclear fission
demonstrates the corresponding instability of insulating 
drops. Appendix \ref{secground} presents several 
mechanisms by which conducting and insulating drops can 
decrease their energy. For example, we show that a 
conducting charged drop can get rid of its {\it entire} 
electrostatic energy by emitting an infinite number of
infinitesimal droplets.

The mechanisms discussed in Appendix \ref{secground} rely 
on the breakup of the charged drop. Such routes are not 
available to the PA chain which must maintain its 
connectivity. Is the charged PA susceptible to similar
instabilities despite its connectivity? Fig.\ \ref{FigM} 
depicts $R_g^2(T=0.1q_0^2/a)/R_g^2(T=\infty)$ as the 
function of the reduced charge $Q/N^{1/2}$ for different
values of $N$. The curves become steeper with increasing 
$N$ and intersect at $Q/N^{1/2}\approx 1.4q_0$. At the 
intersection point the radii scale as self--avoiding 
walks (with a prefactor slightly larger than the infinite 
temperature value. For $Q>1.4Q_c$ the PAs at low $T$ are 
more stretched than self--avoiding walks, in disagreement
with the finite elongation predicted for ellipsoidal shapes.
How does the PA go beyond the ellipsoidal limit while 
maintaining its connectivity? We can still exploit analogies 
to charged drops.

The {\it annealed} PA is modeled by a deformable 
{\it conducting} ``drop'' of particles of size $a$, 
constrained to maintain a spherical  topology. Although such 
a drop cannot expel charged particles, it can still reduce 
its energy by extruding charge in a  finger of length $L$ 
and diameter $a$. Balancing the Coulomb energy ($Q^2/L$) of 
the finger with the increase of surface energy $\gamma a L$, 
we find that the optimal finger length is 
$L\approx Q/\sqrt{\gamma a}$. Fingers appear spontaneously
only if their cost  (roughly $Q\sqrt{\gamma a}$) is less 
than the Coulomb energy of the uniformly charged sphere, 
$Q^2/R$, i.e. for $Q>R\sqrt{\gamma a}$. The fingering 
instability occurs for $\alpha\approx a/R\ll1$, i.e.
far below the shape instability of a sphere. Thus the 
typical annealed PA has a protruding finger of length 
$L\propto Q\propto N^\beta$ for $\beta>1/3$. But, as the weight of
the finger is small, it does not effect the scaling of 
$R_g^2$ for $\beta<5/9$. Such PAs  have large spanning 
sizes without appreciably greater $R_g$.

The {\it insulating} (uniformly charged) drop follows a different 
route. As is shown in Appendix \ref{secground}, it is stable 
for small $\alpha$, but reduces its energy by splitting into 
several droplets of equal size for $\alpha>0.293$. We can 
again constrain the overall object to  remain singly 
connected by linking the droplets via narrow tubes of total  
length $L$ and diameter $a$. As long as $L a^2\ll R^3$, most 
of the charge  remains in the spheres. The total 
electrostatic energy is proportional to $Q^2/L$,
while the surface energy cost grows as $\gamma a L$. Equating 
the two gives $L\propto Q$; not surprisingly, of the same 
order as the fingers in the conducting case. However, 
whereas the surface tension in the conducting case results 
in one big central drop, for the insulating case the droplets 
are separated as in a {\it necklace}. The radius of gyration  is now 
of the same order as the span of the necklace, $L$.

\section{Randomness in The Necklace model}\label{secbeyond}

The {\it necklace model} provides a good picture of a 
polymer with a short--range attractions between its monomers 
and a {\it uniformly} distributed excess charge: A polymer 
with $Q\gg Q_R$ is split into roughly $\alpha\propto Q^2/N$ 
beads connected by a string. Each bead is just below the 
Rayleigh threshold, and the string is stretched by their
Coulomb repulsion to a length $L\propto Q$. (Note that, as 
shown in Appendix \ref{secground}, the optimal number of beads is 
proportional to $(Q/Q_R)^2$ and not $Q/Q_R$.) For 
$Q\ll q_0N$, only an infinitesimal fraction of monomers are 
part of the string, and the overall extensive part of the 
energy is unchanged. This picture should extend to any
deterministic sequence, e.g. composed of alternating
charges\cite{victor}, which has a compact state when uncharged.

Is the necklace model also applicable in the presence of 
random  charges? For our model PAs $\alpha\approx Q^2/(q_0^2N)$,
with a prefactor almost identical to unity, and we shall 
use this relation as an exact definition of $\alpha$. For 
$\alpha\gg1$ we may try to split a chain into $\alpha$ 
segments of approximately equal size. Each segment has 
average charge $Q/\alpha\propto N/Q$ and incorporates 
$N/\alpha\propto (N/Q)^2$ monomers. Thus the fluctuations 
in charge of each segment are of the order of the average 
charge itself, and the picture of uniform, mutually  
repelling, beads is no longer applicable. It is not clear
 how we should model the shapes and distribution of the 
segments which have $\alpha$s of order one.

Let us illustrate the difficulties caused by randomness for 
the case of an unrestricted PA. Since $\overline{Q^2}=q_0^2N$, 
where the overline denotes an average over the ensemble of 
all quenches, we have $\overline{\alpha}=1$. As  demonstrated
in Appendix\ \ref{secground}, the insulating drop is unstable 
to splitting already for $\alpha\approx0.3$, and thus a 
typical random PA is expected to form several globules 
connected by narrow tubes. Now consider splitting  sequences 
of $N$ monomers with total charge constrained to a particular 
$\alpha$ into two equal subchains of charges $Q_1$ and $Q_2$. 
It is easy to show that each segment has
$\overline{\alpha_{\rm subchain}}=(1+\alpha)/2$, while the mean  
product of the charges is $\overline{Q_1Q_2}=q_0^2N(\alpha-1)/4$. 
The subchains have, on average, values of $\alpha$ close to 
unity. Also, for $\alpha=1$, the  average value of the product 
of charges vanishes. We thus have the paradoxical situation in 
which most spherical shapes are unstable, while there is on 
average no energetic gain in splitting the sphere into two 
parts. It is most likely that the ensemble of chains with 
$\alpha\approx 1$ contains a broad distribution of sizes and 
shapes.

Thus charge inhomogeneities drastically modify the necklace  
picture. The resulting PA is probably  still composed of rather 
compact globules connected by a (not necessarily linear) 
network of tubes. The globules are selected preferentially from 
segments of the chain that are approximately neutral (or at 
least below the instability threshold), while the tubes are 
from subsequences with larger than average excess charge. It is 
amusing to inquire how a random sequence is best partitioned 
into large neutral segments. The resulting segments appear
to have a broad distribution which will be addressed in future publication\cite{KKfuture}.

In summary, we find that the behavior of PAs, and other charged  
polymers is controlled by the parameter $\alpha\propto Q^2/N$. 
Chains with small values of $\alpha$ form compact spherical 
globules. The globules split for $\alpha>0.3$, resulting in a 
necklace of beads if the charge inhomogeneity is small. The 
span of the uniform necklace scales with the net charge $Q$. 
We don't have a consistent theoretical picture for the random 
PA beyond the instability threshold. The numerical results in 
Fig. \ref{FigM} suggest that the size of such PAs grows faster 
than that of a self--avoiding walk, i.e. $\nu>0.6$. The
simulations so far are not inconsistent with $\nu=1$ suggested 
by a scaling argument\cite{rKK}. However, as the simulations 
suffer from the usual shortcomings of small sizes, sampling,
and equilibration, a definitive answer about the behavior of 
PAs is still lacking.

\acknowledgments
We would like to thank B.I. Halperin for bringing the importance 
of surface tension to our attention, and D. Erta\c s for helpful  
discussions. This work was supported by the US--Israel BSF grant 
No. 92--00026, by the NSF through No. DMR--87--19217 (at MIT's 
CMSE), DMR 91--15491 (at Harvard), and the PYI program (MK).

\appendix
\section{Monte Carlo Procedure}  \label{MC}

The model used for Monte Carlo (MC) simulations is described 
in the Section \ref{secTdep}. Discretizing monomer locations 
simplifies checking for excluded volume interactions; allowing 
the bond length between the nearest neighbors to fluctuate 
without energetic cost  (a ``square--well'' potential) provides 
sufficient flexibility to facilitate equilibration. The 
square--well potential has been used before in continuum 
simulations of tethered surfaces\cite{rKKN}; on discrete 
lattices it is known as the {\it fluctuating bond 
method}\cite{rCK}. The details of the MC procedure are as 
follows.

For each $T$ and $Q$ we start by selecting a (quenched) random 
sequence of charges $\pm q_0$, whose sum (the total excess 
charge) is fixed to $Q$. This is accomplished by randomly 
selecting $(N-Q/q_0)/2$ positions on the chain for negative 
charges, and placing positive charges on the remainder. For 
each quench, we perform a thermal equilibration at temperature 
$T$, and then calculate the thermal averages of interest.
The thermalization is repeated for ten different sequences
and the results are averaged over the quenches. In an 
elementary MC step a monomer is picked at random and moved
a single lattice unit. The move is accepted according to the 
usual Metropolis rule. Since each move requires recalculation 
of interaction energies, it involves $O(N)$ operations. The MC 
time unit is defined as the period during which $N$ attempts 
are made. Thus, the CPU time per single MC time unit increases 
as $N^2$.

Obtaining good averages in random systems is a significant 
challenge. Errors appear due to both inadequate thermal 
equilibration and insufficient quench averaging.  Our 
high--temperature chains (for $T>5q_0^2/a$) are similar to 
uncharged polymers and their equilibration is limited by 
the slowly decaying ``Rouse modes". The slowest decay time
is approximately the interval taken by a polymer to diffuse its 
own radius of gyration, estimated as follows: Since the 
acceptance rate of an elementary MC move is of order one 
throughout the simulation, the diffusion constant of a 
single monomer is also of order one (in units of squared 
lattice constant divided by the MC time unit). The
diffusivity of the polymer center of mass is $N$ times 
slower, resulting in a  diffusion constant of 
$D\approx a^2/N$, and a relaxation time of 
$\tau'=R_g^2N/a^2$. At high temperatures $\tau'$ scales as
$N^{1+2\nu}$, where $\nu=0.588$. At low temperatures 
($T<0.1q_0^2/a$) the polymer is almost compact, and a 
characteristic time can be obtained by considering  phonons,
plasma oscillations, or large--scale density fluctuations. 
Such time scales, in our MC time units, grow as 
$\tau''\approx R_g^2/a^2\sim N^{2/3}$. Unfortunately, there 
are probably much slower (and more important) time scales
associated with crossing over large barriers to shape  
rearrangement which are thermally activated. We have no 
estimates for such times.

We used $\tau\equiv N^2$ MC units as the basic equilibration 
time. Each equilibration lasted $250\tau$, but the first 
$10\tau$ configurations were dismissed in calculating thermal 
averages. As the number of operations per equilibration 
increases as $N^4$, this is close to the maximal equilibration 
time which can be reasonably used in a simulation of this type. 
Several hours of CPU time (on Silicon Graphics R4000 
workstation) were needed to equilibrate each quench at a given 
temperature for $N=64$. Consequently, more than a day of CPU 
time is used to generate a single date point by averaging over 
10 quenches. To collect all the data on $N=64$ chains we needed 
about two months of CPU. For $N=128$ we spent 10 days of CPU to 
obtain a single data point, and therefore only the $Q$ 
dependence at a single temperature was investigated.

We believe that the times used in equilibration produce 
satisfactory thermal averages. A direct check of the temporal 
correlation function of the radius of gyration for $N=64$ 
indeed indicates that the correlation time is approximately 
equal to $\tau$ at high temperatures. This suffices to produce 
very good thermal averages. For example, a particular sequence 
of $N=64$ monomers with $Q=0$ at $T=25$ has average 
$R_g^2=152a^2$, with standard deviation of approximately 
$60a^2$. For this polymer $\tau'=152*64=2.5\tau$, and thus our 
simulation contains approximately 100 independent 
configurations. Therefore the average value of $R_g^2$ is 
accurate to about $\pm6a^2$. However, the average $R_g^2$ for 
10 distinct quenches are scattered over an interval of width 
40$a^2$. Thus the accuracy of our thermal averaging suffices to 
show that different quenches have slightly different thermal 
averages of $R_g^2$. The accuracy of the final average is, 
therefore, limited by the number of quenches rather than by 
the thermal averaging.

The  correlation time obtained at low temperatures from 
temporal correlation functions of neutral PAs is shorter.
This just reflects the reduction in size of the entire 
polymer. As mentioned earlier, correlations of   $R_g$ are 
rather insensitive to shape changes and their much longer 
activated time scales. To obtain some, admittedly indirect, 
measure of the quality of equilibration for dense polymers, 
we compare our simulations with the quite extensively
investigated  {\it restricted primitive model} (RPM).
The latter represents a solution of positive and negatively 
charged particles ($\pm q_0$),  interacting via a  Coulomb 
force and a hard core repulsive potential of diameter 
$\sigma$. (For a review of the subject see Ref.\cite{Fisher}.)
The thermodynamics of the model is conveniently presented
in terms of a dimensionless density $\rho_*\equiv n\sigma^3$ 
where  $n$ is the actual number density, and a dimensionless 
temperature $T_*\equiv k_BT\sigma/q_0^2$. At low temperatures 
the solution undergoes a phase separation between high and low  
density phases. As indicated by the dashed line in 
Fig.\ \ref{FigN},  the critical point occurs at very low 
density (unlike regular fluids with short range interactions).
Numerical investigations of this phase  
transition\cite{FishLev,Friedman,Valleau,Pana}
have encountered considerable difficulties: despite the low 
density, the behavior of the system becomes very erratic close 
to the critical point.

Since our simulations also involve low temperatures and 
relatively high densities, it is interesting to find out where 
our system is located on the $(T_*,\rho_*)$-plane. Of course, 
as we are dealing with a single polymer rather than a dense 
solution, the comparison involves few somewhat arbitrary factors. 
Our lattice potentials approximately mimic interactions of the 
hard core particles of RPM. From this comparison we relate the 
MC temperature to $T_*$ by $T_*=1.2Ta/q_0^2$. Secondly,
we calculate the polymer density, assuming that the monomers  
uniformly occupy the volume of a homogeneous ellipsoid with 
identical eigenvalues of the shape tensor $\{\lambda_i\}$.
This leads to  a reduced density, 
$\rho_*=2.4a^3/\sqrt{\lambda_1\lambda_2\lambda_3}$.
For a neutral PA, $\rho_*$  is approximately independent of $T$ 
for $T>5q_0^2/a$, and increases at lower $T$, leading to the 
trajectory indicated by the  solid line in Fig. \ref{FigN}. 
At densities close to the critical density of RPM, the PA
temperature is almost an order of magnitude higher than the 
critical temperature. At lower temperatures the trajectory of 
PA simulations approaches the phase boundary on the ``liquid''
side, at densities twice higher than the critical density
of RPM. Thus our polymers stay away from the problematic
region where critical fluctuations may cause significant
equilibration problems. Since our equilibration times exceed 
by several orders of magnitude those used in RPM model 
simulations, we believe that we have well equilibrated results.

The acceptance rate of MC moves in our simulations is 
approximately 0.6 for all temperatures. It drops to 0.46 at  
$T=0.05q_0^2/a$, and further lowering of temperature leads to 
a gradual ``freezing''. Repeated heating and cooling cycles 
performed on several samples indicates that the behavior is 
essentially reversible for $T>0.05q_0^2/a$. We believe that 
the configurations obtained for $T=0.05q_0^2/a$ are very close
to the actual ground state. Most of the low temperature  
investigations were actually performed at $T=0.1q_0^2/a$ where 
the thermal averages are more reliable. For chain length $N=64$ 
we performed an extensive study of the dependence of $R_g$ and 
other quantities on $T$ and $Q$. We also investigated the 
$N$--dependence of these quantities at $T=0.1q_0^2/a$ for 
$N=16$, 32, 64, and 128.

\section{Spheroidal Distortions of Charged Drops} \label{secchdrop}

In this Appendix we discuss the minimum energy shape of a 
single charged drop. We first examine the conducting drop, 
and then relate the results to insulating ones in the last 
paragraph. In terms of the dimensionless parameter $\alpha$,
Eq.\ (\ref{Esphere}) for the energy of a charged conducting   
spherical drop is
\begin{equation}\label{Ealpha}
E(Q)=E(0)(1+2\alpha).
\end{equation}
For large $Q$ the spherical shape is unstable, and the
stable shape is determined by minimizing the sum of surface 
and electrostatic energies. In an equilibrium shape the 
surface is an equipotential, since otherwise energy can be 
reduced by redistributing the surface charge. The pressure 
difference between the inside and outside of the drop at 
any point on the surface is given by, 
\begin{equation}
\label{edp}
\Delta p=\gamma({1\over r_1}+{1\over r_2})-2\pi\sigma^2\ ,
\end{equation}
where $r_1$ and $r_2$ are the principal radii of curvature,
and $\sigma$ is the surface charge density. In equilibrium 
the pressure inside the drop must be constant. Thus, 
Eq.\ (\ref{edp}), with a constant value of $\Delta p$ at all 
points, determines the equilibrium shape. This is in fact a  
rather complicated integro--differential equation whose 
general solutions are not known. Note that for $Q=Q_R$, we 
find $\Delta p=0$ for a spherical drop.

In 1882 Lord Rayleigh investigated the stability of a charged
drop\cite{Rayleigh} and showed that for $\alpha=1$ the sphere 
becomes unstable to surface distortions described by the 
Legendre function $P_2(\cos\theta)$. (The points of 
instability for higher harmonics are given by
$\alpha_n=(n+2)/4$.)
The instability does not result in a small distortion  as
$\alpha$ exceeds unity but, rather, leads to a strongly 
elongated shape. For $\alpha>1$ no exact analytical treatment 
is available. Some progress is possible by assuming that the 
drop is an ellipsoid of revolution (i.e. a prolate 
spheroid)\cite{Taylor}. Since both the surface
area and the electrostatic energy of such shapes are known
(see, e.g., Ref.\cite{Jeans}), the problem reduces to the  
minimization of
\begin{eqnarray}
\label{eQ}
E(Q)&=&{E(0)\over 2}\left[(1-e^2)^{1/3}\left(1+{\sin^{-1}e\over
e\sqrt{1-e^2}}\right) \right. \nonumber \\
 &+&\left. 2\alpha{(1-e^2)^{1/3}\over  
e}\ln\left((1+e)/(1-e)\right)\right]
\end{eqnarray}
with respect to eccentricity $e\equiv\sqrt{1-b^2/a^2}$, where 
$a$ and $b$ are the major and minor semi-axes. The first term 
in the brackets is the total surface area (incorporating the 
constraint of fixed volume), while the second term is the 
electrostatic energy ($Q$ enters via $\alpha$). It has been 
shown\cite{Taylor} that a prolate spheroid is {\it not} an
exact equilibrium shape: for aspect ratio $a:b=2:1$ the pressure  
difference $\Delta p$ in Eq.\ (\ref{edp}) varies by about 2\% 
for different points on the surface. Nevertheless, we may 
assume that the forces pulling the ellipsoid out of the shape 
are small as long as $a/b$ is not excessively large. Numerical 
solutions of a drop in an external field confirm that the 
ellipsoidal approximation is reasonably good for shapes that  
are not too elongated\cite{Miksis,Sero,Li}.

\begin{figure}
\caption{Energy differences between drops of prolate spheroidal, 
and spherical, shape (normalized to the energy of spherical drop) 
as a function of squared eccentricity $e^2$. The graphs correspond 
(from top to bottom) to $\alpha=0.88$, 0.89, 0.895, 0.898, 0.90, 
0.91, and 0.92.}
\label{FigK}
\end{figure}
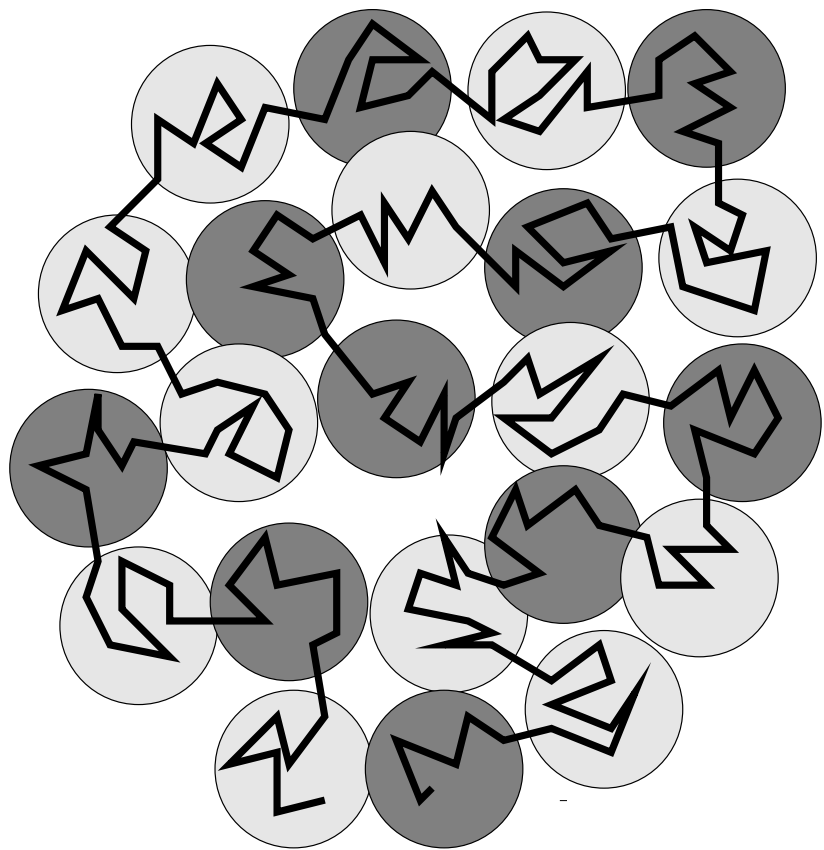
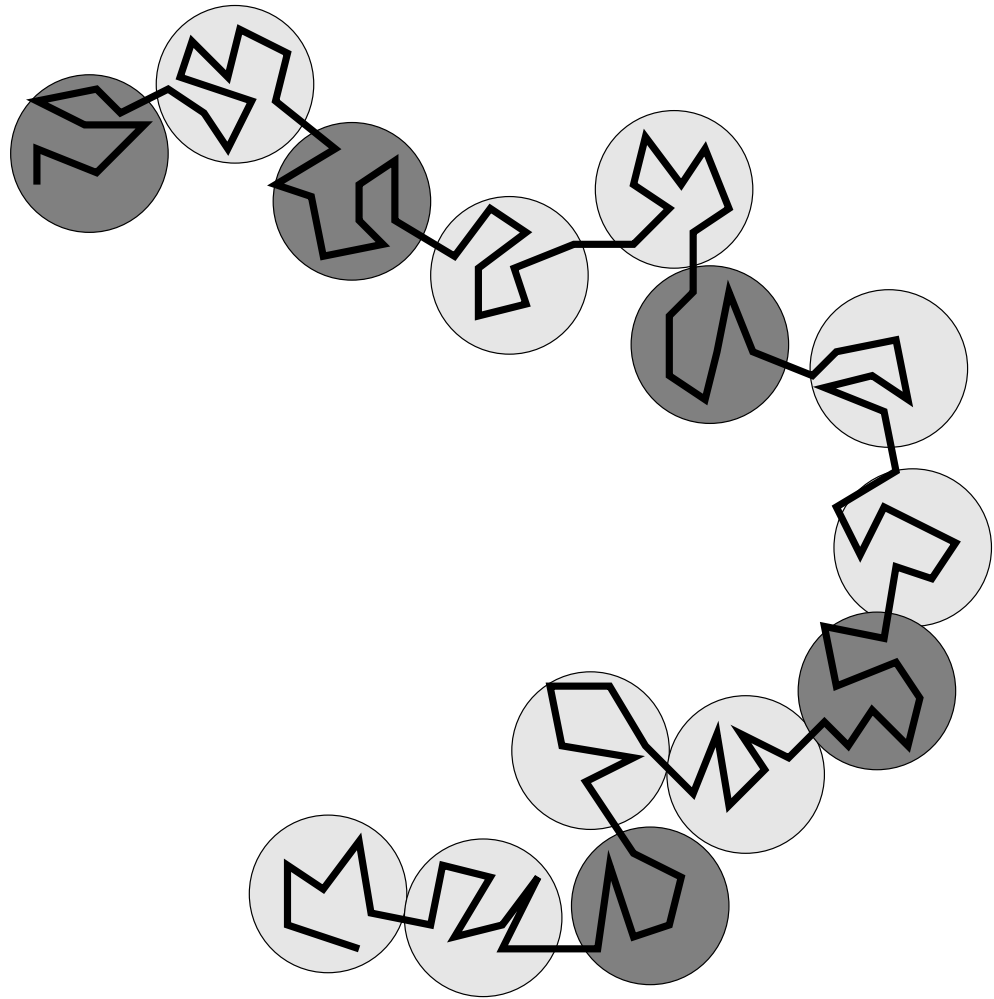
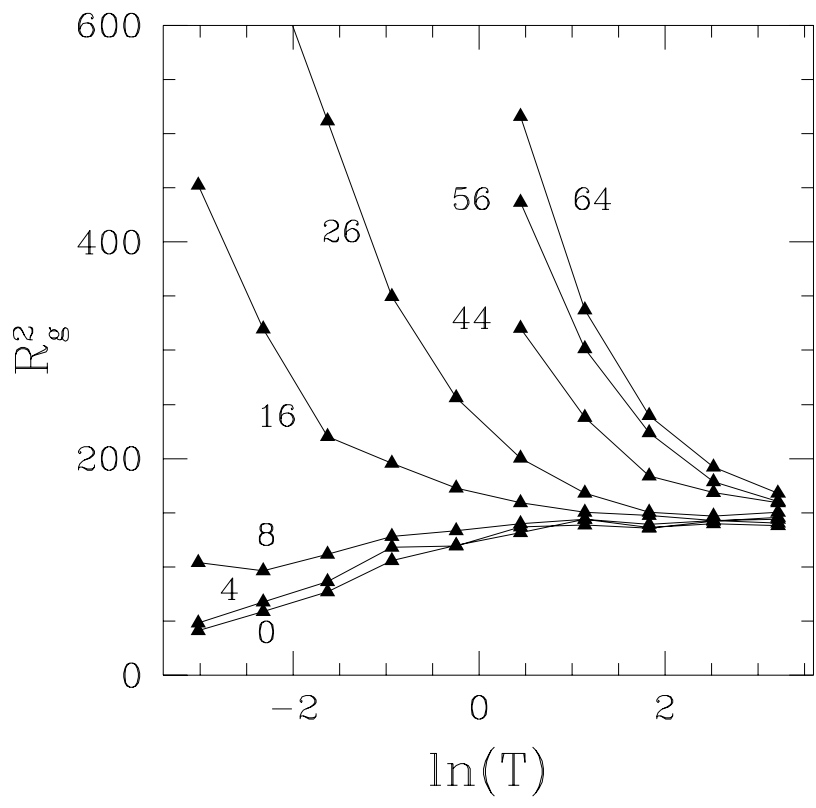
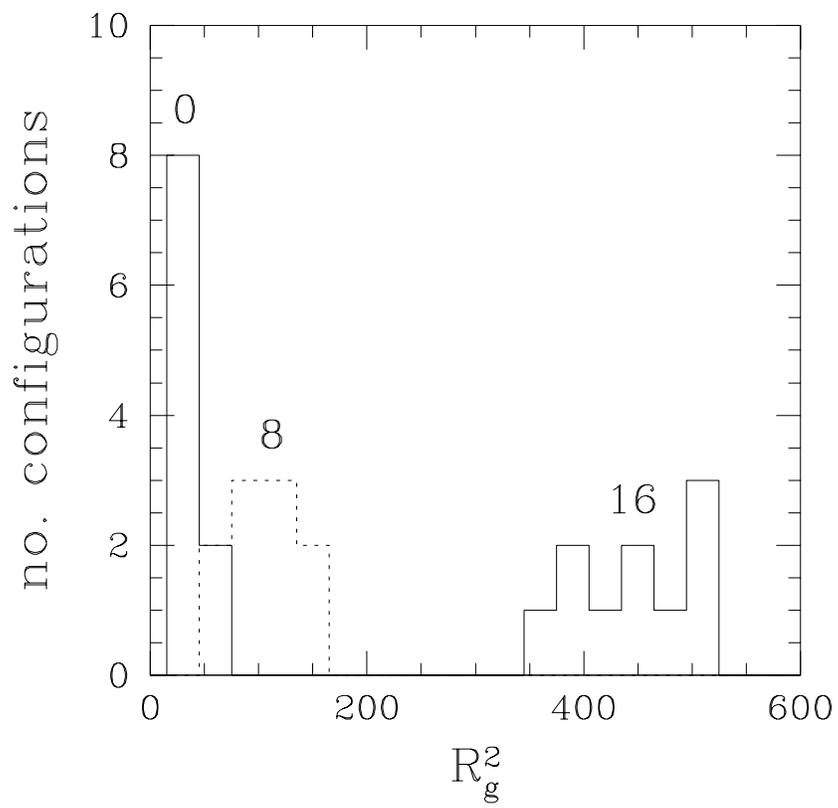
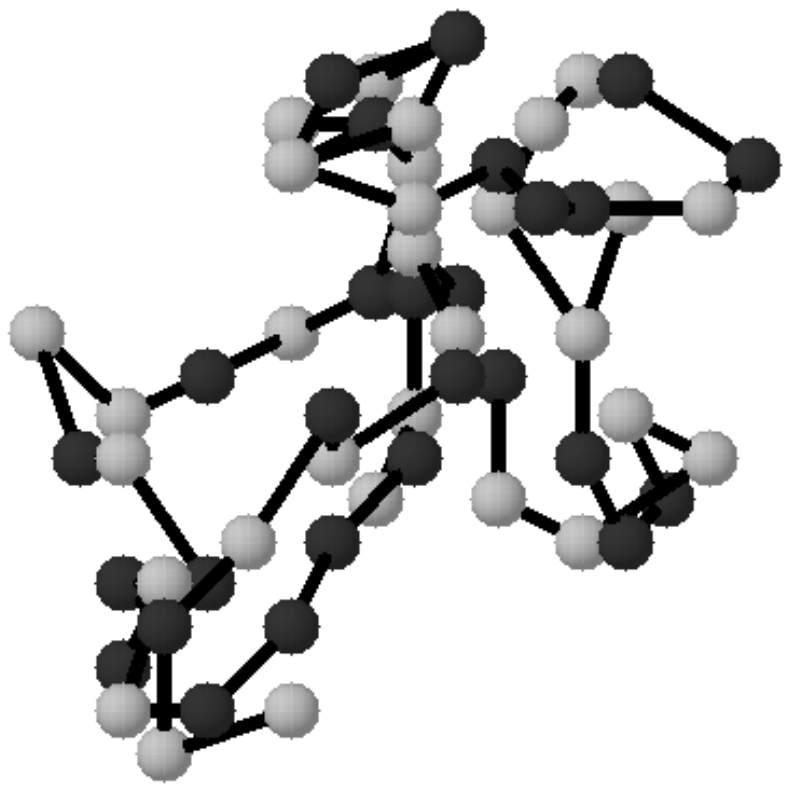
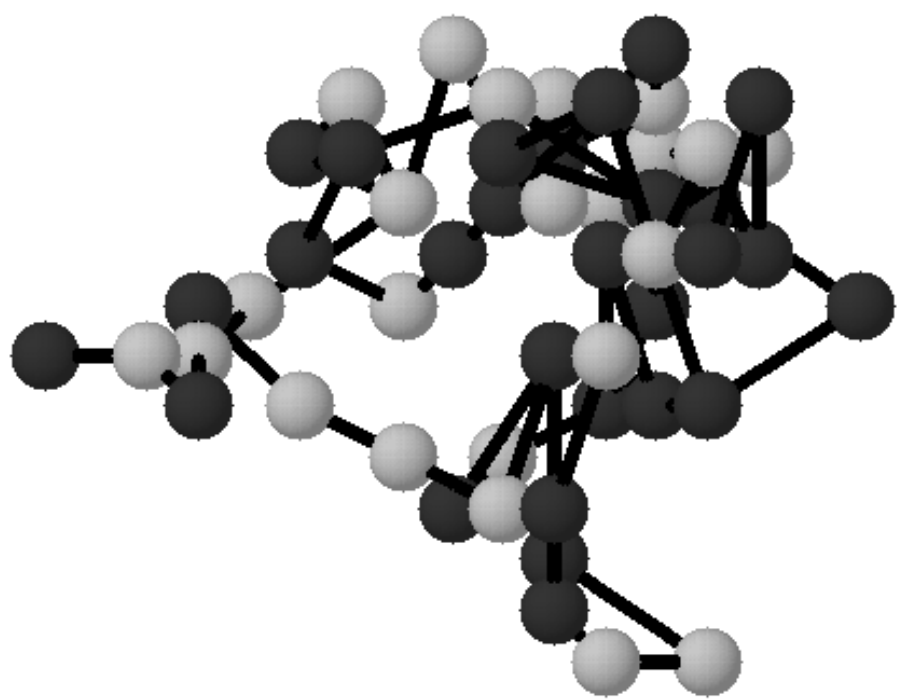
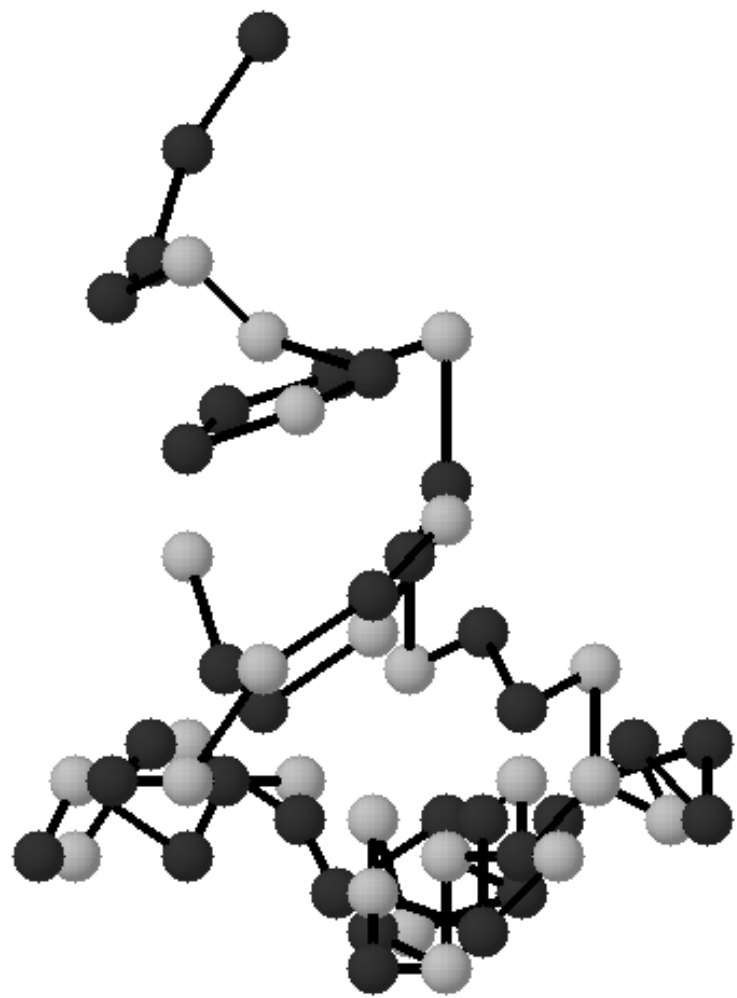
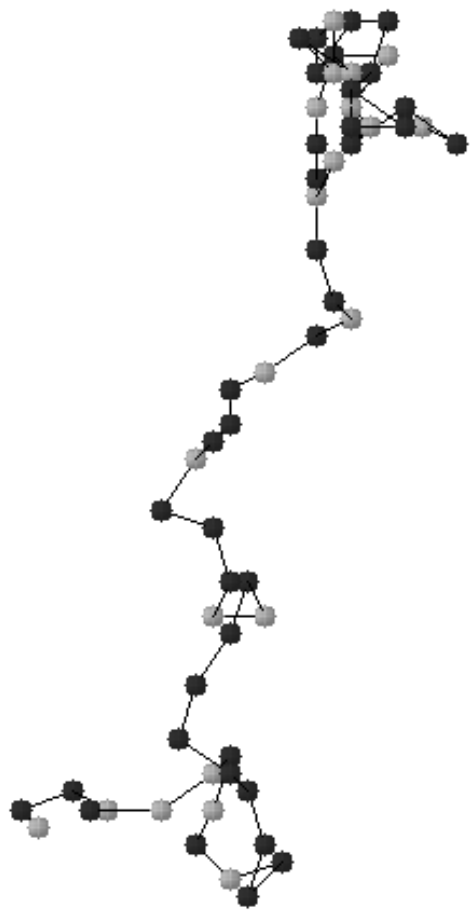
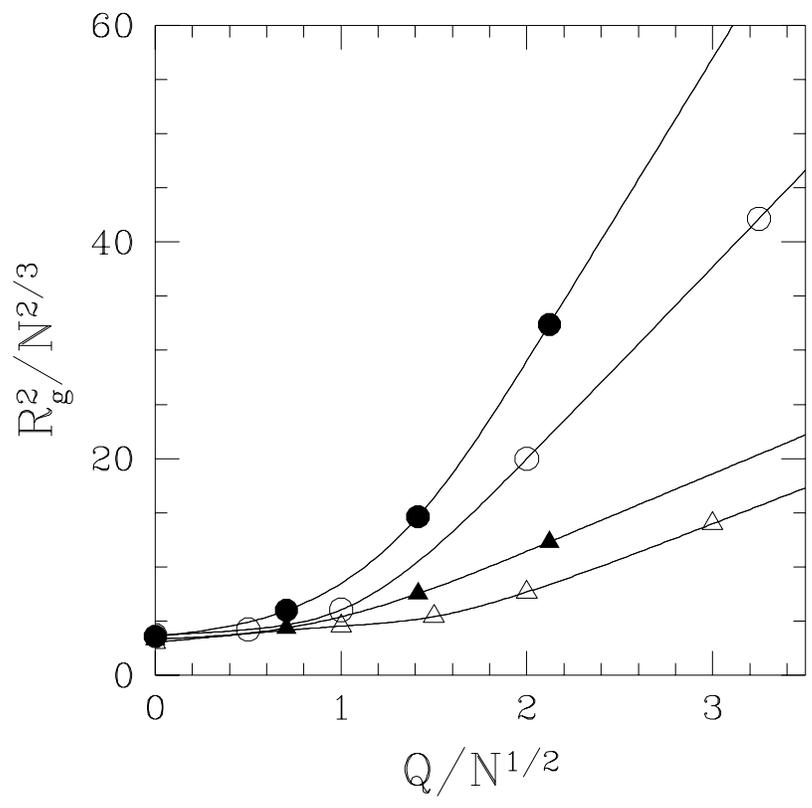
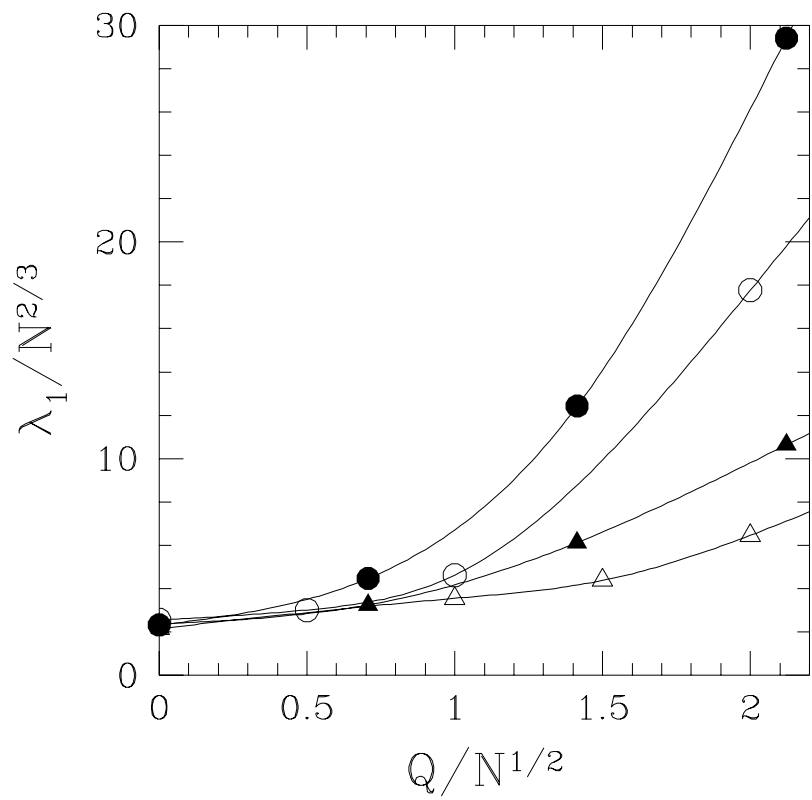
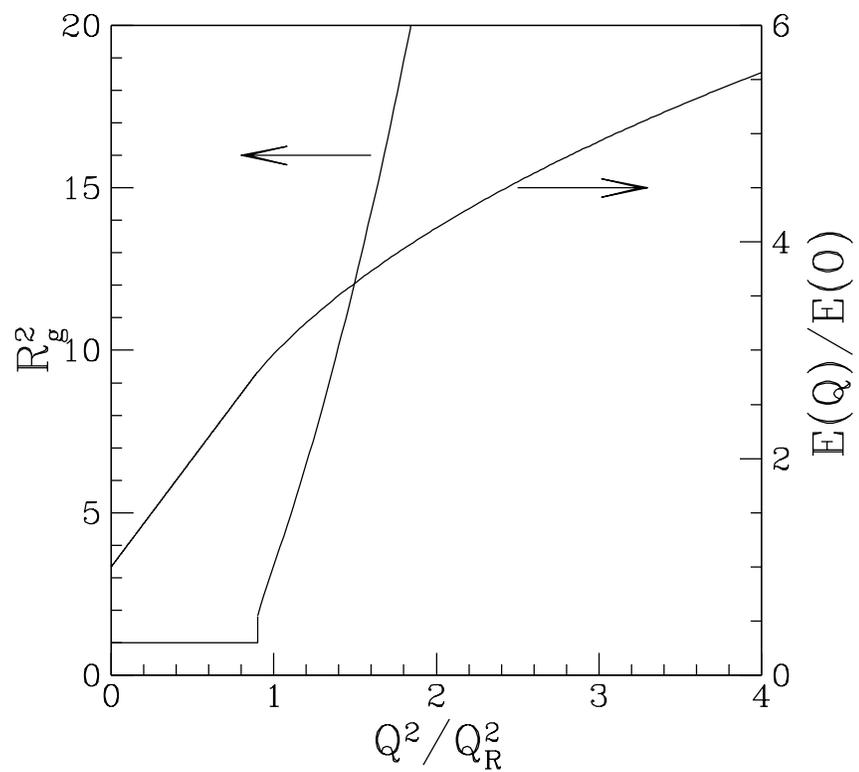
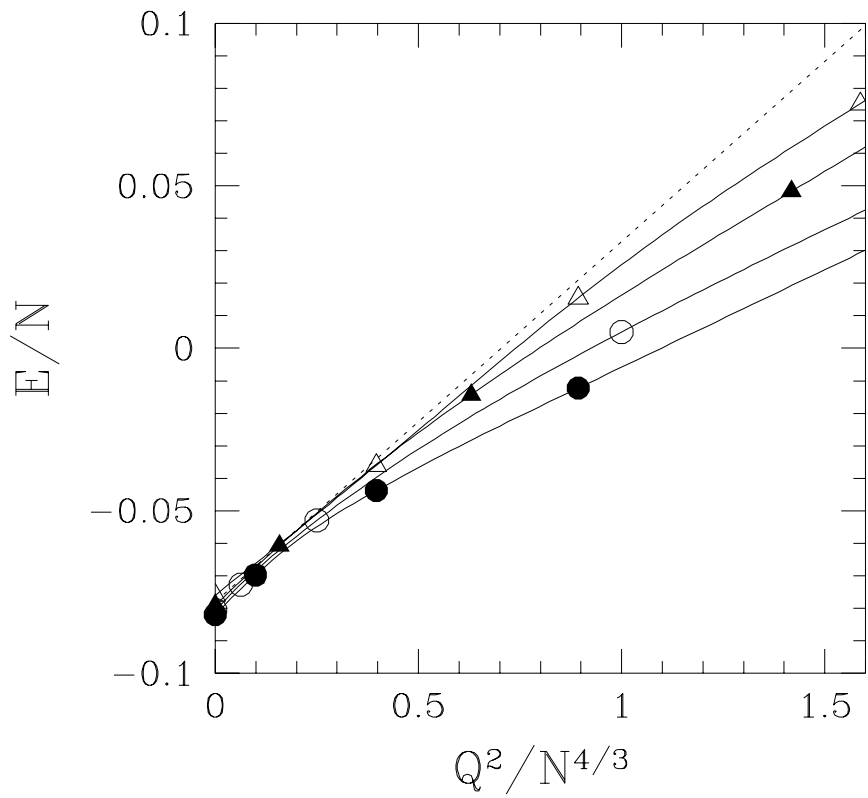
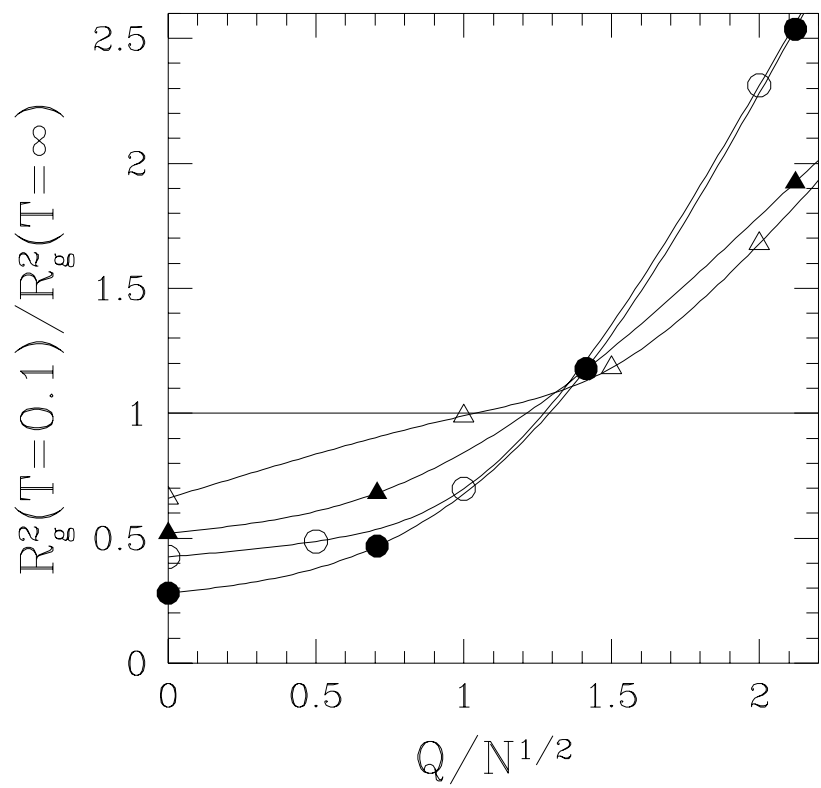
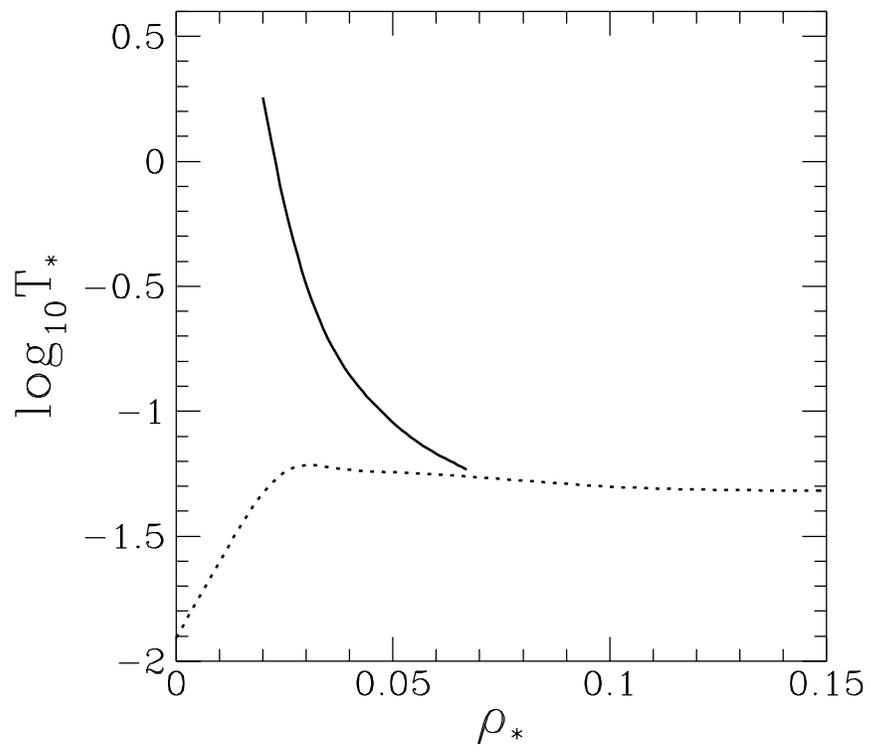
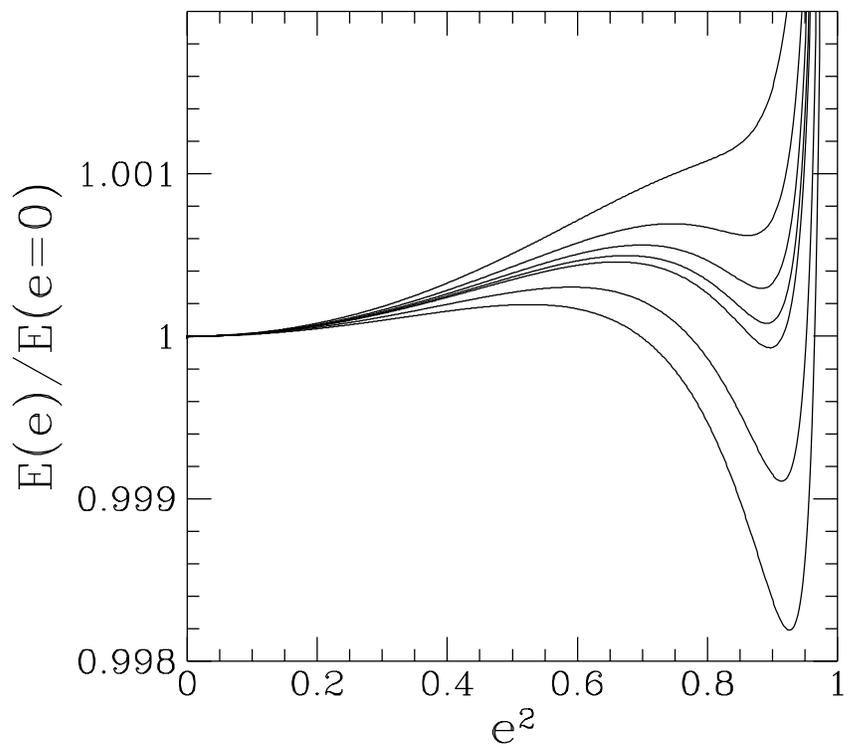

Ailam and Gallily\cite{Ailam} noted that Eq.\ (\ref{eQ}) has a 
local minimum for $e \ne 0$ even for $\alpha$ smaller than 
unity, i.e. below the Rayleigh stability limit. However, they 
did not determine this range accurately. Fig. \ref{FigK} 
depicts $E(Q)$ as a function of $e$ for several values of 
$\alpha$ in the relevant parameter range. A new local minimum 
first appears for $\alpha=0.887$, and becomes the {\it global} 
minimum for  $\alpha>0.899$. At the latter  $\alpha$ a 
spherical drop should discontinuously ``jump'' to a strongly  
elongated shape with $e=0.95$. However, the spherical shape 
($e=0$) remains a {\it local} minimum until $\alpha=1$. In an 
ideal experiment in which the charge is gradually increased 
the drop stays in a metastable spherical shape. At $\alpha=1$, 
the sphere becomes unstable and stretches to $e=0.98$. For 
$\alpha\gg1$, the eccentricity approaches unity. In this 
limit, the asymptotic forms of the surface and electrostatic  
energy are $(1-e)^{-1/6}$ and $-(1-e)^{1/3}\ln(1-e)$ 
respectively. The minimum energy is achieved at
$e^2\approx 1-(\pi/8\alpha)^2/\ln^2(\pi/8\alpha)^2$, while the
energy increases as
\begin{equation}
\label{eellips}
E(Q)\approx 2.03 E(0)(\alpha\ln\alpha)^{1/3} \ .
\end{equation}
Note that the asymptotic increase of $E(Q)$ ($\propto Q^{2/3}$)  
is much smaller than that of the undeformed sphere ($\propto Q^2$). 
The exact dependence of $E(Q)$ on $\alpha$ is plotted in 
Fig.\ \ref{FigL}. In the same figure we also present the squared
radius of gyration of the elliptical drop,
\begin{equation}
{R_g^2(Q)\over R_g^2(0)}={a^2+2b^2\over 3R^2}=
{(1-e^2)^{-{2\over3}}+2(1-e^2)^{1\over3}\over3}.
\end{equation}
Since for large $Q$,  $1-e^2$ is proportional to $\alpha^{-2}$ 
up to logarithmic terms,  
$R_g\sim R\alpha^{2/3}\sim R(Q/Q_R)^{4/3}$.
While the ellipsoidal shape is not an exact solution to the 
charged drop problem, Eq.\ (\ref{eellips}) provides an upper 
bound on the total energy.

We next consider a drop in which the charge is {\it uniformly}  
distributed over the volume. This system has been considered 
in the context of the liquid drop model of atomic nuclei (see,
e.g., Ref.\cite{weisskopf}). The energy of  the uniformly 
charged sphere, Eq.\ (\ref{EiQ}), is
\begin{equation}\label{Einsul}
E^{(i)}(Q)=E(0)({12\over5}\alpha+1)\ ,
\end{equation}
As in the case of the conducting drop, the uniformly charged 
drop becomes locally unstable\cite{bohr,feenberg,weizsacker} 
to infinitesimal distortions for $\alpha=5/6$. Assuming that 
the drop distorts into a prolate spheroid, its energy can be 
written down explicitly\cite{weizsacker}.  The resulting 
energy is identical to Eq.\ (\ref{eQ}), except that the factor 
of $2\alpha$ in the second term  is replaced by ${12}\alpha/5$.
The same factor relates Eqs. (\ref{Ealpha}) and (\ref{Einsul}).
Thus all the previous results for  conducting drops are also
applicable to insulating drop after multiplying $\alpha$s by 
$6/5$. We should note that this simple substitution does not
hold for drops of arbitrary shape.

\section{Splitting a Charged Drop} \label{secground}

An elongated ellipse is not a {\it local} equilibrium shape of a  
drop, since perturbations $P_n(\cos\theta)$ with $n>2$ become unstable\cite{Grig}. There are other theoretical indications 
that no elongated shape is the {\it global} ground state. For 
example, as the eccentricity increases the electric field at
a tip ($\sim Q/b^2\sim Q(1-e^2)^{-1/3}$) becomes strong enough
to support conical tips\cite{Taylor,Li}. Experimentally, it 
is observed that, for $\alpha>1$, a conducting charged
drop disintegrates into smaller ones. Assuming that the 
experiments can be described by the charged drop
model\cite{Zeleny,Elghazaly,Ryce},  disintegration
begins for $\alpha$ equal to or slightly  above one.
This is not surprising, since already at $\alpha=1$ the 
equilibrium shape is strongly deformed. The instabilities 
lead to ejection of smaller droplets whose size distribution 
is believed to be controlled by hydrodynamic effects. 
Apparently, after a significant elongation, many roads 
towards decreasing energy open up and the choice is made by 
dynamical effects. In this Appendix we shall investigate the 
splittings of a drop into several spherical droplets removed 
to infinite separations.

First consider the splitting of a single {\it conducting} 
drop of radius $R$ into two (secondary) droplets with radii 
$R_1$ and $R_2$. Using Eq.\ (\ref{Ealpha}) for the energy of 
each droplet, the total energy of the infinitely separated 
pair is
\begin{equation}\label{Epair}
E_2(Q)={q_1^2\over 2R_1}+{q_2^2\over 2R_2}
+4\pi\gamma(R_1^2+R_2^2)\  . 
\end{equation}
The charges satisfy $q_1+q_2=Q$, while the radii are 
constrained by $R_1^3+R_2^3=R^3$ to preserve the total volume.
Minimizing $E_2(Q)$ with respect to $q_1$, while treating 
$q_2$ as a dependent variable, gives 
\begin{equation}\label{qviar}
q_i={R_i\over \sum_j R_j}Q
\end{equation}
Substituting this result into Eq.\ (\ref{Epair}) we find
\begin{equation}\label{Epairreduced}
E_2(Q)=E(0)\left({2\alpha\over \sum_j r_j}+\sum_j r_j^2\right)\ ,
\end{equation}
where the reduced radii $r_j\equiv R_j/R$ satisfy the fixed volume
constraint $\sum_j r_j^3=1$. (Note that $E(0)$ denotes the surface
energy of the {\it original} drop.) The stationary value of
$E_2(Q)$ is found by solving $\partial E_2(Q)/\partial r_1=0$,
and treating $r_2$ as a dependent variable. The resulting
equation has several possible solutions, including
\begin{equation}
r_1^3=r_2^3={1\over2}.
\end{equation}
However, the symmetrical solution is a {\it minimum} only for  
$\alpha\ge 1$. Thus strongly charged drops would prefer 
splitting into two equal droplets. A second solution,
\begin{equation}
r^3_{1,2}={1\over2}\left(1\pm\sqrt{1-{4\alpha^3\over  
1+3\alpha}}\right),
\end{equation}
exists only for $\alpha\le1$, where it is an energy minimum, less
than that of a single sphere. Thus a weakly charged drop can always
reduce its energy by splitting into two unequal parts.

Further breakup of the drop is possible, and we next consider  
splittings into $n$ droplets. The solution to this problem is 
analogous to the previous case. The distribution of charge 
among the droplets is still given by Eq.\ (\ref{qviar}) and 
$E_n(Q)$ has exactly the same form as Eq. (\ref{Epairreduced}) 
with $j$ summed from 1 to $n$. The search for extrema of $E_n(Q)$ 
leads to several solutions, classified by two sets of droplets. 
One set contains $m$ small droplets of reduced radius $a$, while 
the remaining $m-n$  have larger reduced radii $b$ ($b>a$), such 
that
\begin{eqnarray}\label{ndrop}
&& ma^3+(n-m)b^3=1\ , \nonumber \\
&& a={\alpha b\over b\left(ma+(n-m)b\right)^2-\alpha}\ .
\end{eqnarray}
The solution for a particular $m$ exists only in a finite range of  
$\alpha$s, and finding which solution represents the global energy 
minimum is quite cumbersome. One limit, however, is easily 
examined: Consider splitting the drop into one large droplet and 
$m=n-1$ smaller drops. In the limit of $n\rightarrow\infty$,
$a^3\approx \alpha/n^2$, and $b\approx1$; the total volume, area, 
and electrostatic energy of the charged droplets, vanishes, while 
they, nevertheless, carry all the charge of the system. Thus the 
energy of the system is reduced to $E(0)$, i.e. the energy of the 
uncharged original drop!

We next consider the insulating {\it uniformly charged} drop. 
If such a drop is split in two, the charge of each droplet will
be proportional to its volume, i.e. $q_i=QR_i^3/R^3$.
Using the energy of a single sphere in Eq.\ (\ref{Einsul}),
the total energy of a pair of infinitely separated spherical  
droplets is obtained as
\begin{eqnarray}
E_2^{(i)}(Q)&=&{3\over5}\sum_j{q_j^2\over  
R_j}+4\pi\gamma\sum_jR_j^2\nonumber\\
&=&E(0)\left[{12\over5}\alpha\sum_jr_j^5+\sum_jr_j^2\right]\ .
\end{eqnarray}
The charges $\{q_j\}$ and the reduced radii $\{r_j\}$ satisfy 
the same restrictions as in the conducting drop. For 
$\alpha<{1/6}$ the only extremum is a maximum at 
$r_1^3=r_2^3={1/2}$. For $\alpha>{1/6}$, this point is a local 
minimum. However, only for $\alpha>0.293$ is the resulting
energy lower than that of  the original drop.  A similar scenario
is found for splitting the drop into $n$ secondary droplets.
Larger values of $\alpha$ are needed to stabilize solutions with
higher $n$, and there is an optimal number of droplets (all of 
the same radius) for each $\alpha$. Since the energy of an 
$n$--drop system is,
\begin{eqnarray}
E_n^{(i)}&=&n\left[{3\over5}\left({Q\over n}\right)^2\left({n\over
R^3}\right)^{1/3}
+4\pi\gamma \left({R^3\over n}\right)^{2/3}\right]
\nonumber \\
&=&E(0)\left[{12\over5}\alpha n^{-2/3}+n^{1/3}\right]\ ,
\end{eqnarray}
the optimal $n$ (for large $n$) is  found from
$\partial E_n^{(i)}/\partial n=0$ as
\begin{equation}
n={24\over5}\alpha\ .
\end{equation}
The total energy of the optimal configuration grows as
\begin{equation}
E_{\rm optimal}^{(i)}(Q)=2.53E(0)\alpha^{1/3}.
\end{equation}
Thus the uniformly charged drop can not lower its energy as  
drastically as its conducting counterpart. Since we have not 
exhaustively searched for other configurations, the above result 
should be regarded as an upper bound to the ground state energy. 
Note that this bound has the same scaling (up to logarithmic 
corrections)  on $\alpha$ as  that of the highly elongated
spheroid.

\begin{figure}
\caption{Qualitative views of the spatial arrangement of `blobs' in
a PA with quenched randomness. Electrostatic interactions within each
blob are smaller than $k_BT$. Lighter and darker shades of the  
spheres
denote predominantly positively or negatively charged blobs.
The DH view assumes that the blobs can rearrange in a pattern
(a) where interactions are screened on long distances. According
to an RG--inspired picture the blobs form a self--similar pattern (b)
with the same interaction energy on all length scales.}
\label{FigO}
\end{figure}
\begin{figure}
\caption{$R_g^2$ (in units of $a^2$)
as a function of $T$ (in units of $q_0^2/a$) for several 
values of the excess charge $Q$ for a 64--monomer chain.
Each point is an average over the 10 independent quenches used 
at each temperature. The numbers near each curve indicate 
$Q/q_0$.}
\label{FigA}
\end{figure}
\begin{figure}
\caption{Histograms of the distribution of the (ten) values of  
$R_g^2$ (measured for $N=64$ at $T=0.05q_0^2/a$) for several 
charges $Q/q_0$, indicated near the histograms.}
\label{FigB}
\end{figure}
\begin{figure}
\caption{Spatial conformations of 64--monomer PAs at 
$T=0.05q_0^2/a$, for values of $Q/q_0$ equal to (a) 0, (b) 4, 
(c) 8, and (d) 16. Dark and bright shades indicate opposite 
charges. The diameter of each sphere is about 0.4 of the
actual excluded--volume range.}
\label{FigF}
\end{figure}
\begin{figure}
\caption{Scaled $R_g^2$ as a function of  $Q/q_0$ for
chain lengths $N=16$ (open triangles), 32 (full triangles),
64 (open circles), and 128 (full circles).}
\label{FigC}
\end{figure}
\begin{figure}
\caption{Scaled largest eigenvalue, $\lambda_1$, of the shape 
tensor as a function of charge $Q/q_0$ for several chain lengths.
The symbols are the same as in Fig.\ \protect\ref{FigC}.}
\label{FigR}
\end{figure}
\begin{figure}
\caption{Radius of gyration of the minimal energy spheroid in  
units of the radius of gyration of the undistorted sphere (left), 
and the energy of the spheroid scaled to that of the uncharged 
sphere (right), as a function of $\alpha$.}
\label{FigL}
\end{figure}
\begin{figure}
\caption{Energy per monomer in units of $q_0^2/a$ versus scaled
excess charge for several chain lengths at $T=0.05q_0^2/a$. 
The symbols are the same as in Fig.\ \protect\ref{FigC}.
The dashed line is the energy an undistorted PA.}
\label{FigS}
\end{figure}
\begin{figure}
\caption{Ratio between squared radii of gyration at $T=0.1q_0^2/a$  
and $T=\infty$ as a function of scaled excess charge.
The symbols are the same as in Fig.\ \protect\ref{FigC}.}
\label{FigM}
\end{figure}
\begin{figure}
\caption{High and low density fluid (``liquid'' and ``gas'')  
transition line (dotted) in the plane, $(T_*,\rho_*)$, where 
$T_*$ and $\rho_*$ are the reduced temperature and density 
(see text) of the Restricted Primitive Model. The solid line 
indicates the trajectory of the neutral PAs used in our MC 
simulations.}
\label{FigN}
\end{figure}
\end{document}